\documentclass[osajnl,amsmath,amssymb]{revtex4}
\DeclareRobustCommand{\baselinestretch{2}}

\usepackage{graphicx}
\usepackage{dcolumn}
\usepackage{bm}

\begin{document}

\title{Photon number resolving detection using time-multiplexing}

\author{Daryl Achilles}
 \affiliation{Clarendon Laboratory, University of Oxford, Oxford OX1 3PU, UK.}
\author{Christine Silberhorn}
\affiliation{Clarendon Laboratory, University of Oxford, Oxford
OX1 3PU, UK.}
\author{Cezary {\'S}liwa}
 \affiliation{Clarendon Laboratory, University of Oxford, Oxford OX1 3PU, UK.}
\author{Konrad Banaszek}
\affiliation{Clarendon Laboratory, University of Oxford, Oxford
OX1 3PU, UK.}
\author{Ian A. Walmsley}
\affiliation{Clarendon Laboratory, University of Oxford, Oxford
OX1 3PU, UK.}
\author{Michael J. Fitch}
\affiliation{Applied Physics Laboratory, Johns Hopkins University,
Laurel MD 20723-6099}
\author{Bryan  C. Jacobs}
\affiliation{Applied Physics Laboratory, Johns Hopkins University,
Laurel MD 20723-6099}
\author{Todd B. Pittman}
\affiliation{Applied Physics Laboratory, Johns Hopkins University,
Laurel MD 20723-6099}
\author{James D. Franson}
\affiliation{Applied Physics Laboratory, Johns Hopkins University,
Laurel MD 20723-6099}

\date{\today}

\begin{abstract}
Detectors that can resolve photon number are needed in many
quantum information technologies.  In order to be useful in
quantum information processing, such detectors should be simple,
easy to use, and be scalable to resolve any number of photons, as
the application may require great portability such as in quantum
cryptography.  Here we describe the construction of a
time-multiplexed detector, which uses a pair of standard avalanche
photodiodes operated in Geiger mode.  The detection technique is
analysed theoretically and tested experimentally using a pulsed
source of weak coherent light.
\end{abstract}

\maketitle

\section{Introduction}
Detectors that can resolve the number of photons in a pulse of
light have a number of applications in the preparation and
detection on non-classical states of radiation.  For example, a
linear optics approach to quantum computation~\cite{KLM,LOFranson}
requires reliable creation of states with superpositions of up to
two photons.  One method of creating nonclassical states is via
conditional state preparation~\cite{Konrad,Pittman,Loock}, which
relies on the ability to distinguish states of different photon
number.  Photon number-resolving detectors could also enhance the
security of quantum cryptography against certain eavesdropper
attacks~\cite{Hwang,Christine}.

According to the quantum theory of photodetection, the signal
obtained from an ideal noise-free detector has a discrete form
corresponding to the absorption of an integer number of quanta
from the incident radiation. In most practical systems, however,
the granularity of the output signal is concealed by the
imperfections of the detection mechanism. For example, when very
low light levels are detected using avalanche photodiodes (APDs)
operated in the Geiger-mode, the electronic signal can be reliably
converted into a binary message indicating with high fidelity
whether an absorption event has occurred or not. However, the gain
mechanism necessary to bring the initial energy of absorbed
radiation to the macroscopic level saturates already with the
absorption of a single quantum, thus completely masking the
information on exactly how many photons have triggered that event.
A similar problem affects most photomultipliers and APDs operated
in the gain mode~\cite{woodward}, where the excess noise of the
gain mechanism makes discrimination of multi-photon detection
events practically impossible.  Photomultiplier tubes have been
used in the past, however, have been used to attain information
about the photon statics of light pulses~\cite{Kumar}.

Many approaches have been taken to construct detectors with photon
number resolution.   These include several cryogenic devices being
developed for the purpose of resolving photon numbers, including
the VLPC~\cite{yamamoto}, a superconducting bolometer~\cite{NIST},
and a superconducting transimpedance amplifier~\cite{sobolewski}.
There have also been proposals to use coherent absorption of light
in an atomic vapor to enable high efficiency photon number
resolving detection~\cite{Kwiat,Imm}, though no such device has
been experimentally demonstrated to date.  However, the quantum
efficiencies of these available detectors are well below the
efficiency of conventional avalanche photodiodes (APDs), excepting
the VLPC, which has a very high intrinsic quantum efficiency.

Unfortunately APDs do not have the ability to distinguish between
different photon number states.  They do however have quantum
efficiencies as high as $80\%$, they are commercially available,
and are easy to operate.  It is these attributes that make APDs
desirable and prompted the proposal for their use in a fibre-based
detection scheme that allows photon number resolving
detection~\cite{BanaszekWalmsley}.  In this proposal it was shown
that, in principle, it is possible to retain the quantum
efficiencies of the APD while adding the ability to resolve photon
number.  Losses critically affect this ability, since they remove
photons randomly from the input state.  Clearly losses in the
optical system will also affect the fidelity of conditional state
preparation.  However, exploiting prior information, e.g. known
correlations between the signal and idler modes of
downconversion~\cite{PDC}, it is possible to significantly improve
our confidence of the measurements by inductive inference using
Bayes' theorem.

In this paper, we provide a detailed analysis of two recent
experiments \cite{US,THEM} which follow the main concept developed
in \cite{BanaszekWalmsley}, but avoids the need for optical
switches~\cite{czech, czech2}. The basic idea is very similar to
the detector cascade concept~\cite{cascade}, which uses $50/50$
beam splitters to split a pulse into $N$ spatial modes, each of
which is monitored with an APD. Here, instead of splitting the
incident pulse into $N$ spatial modes we split it into $N$
temporal modes, separated by a time interval of $\Delta t$,
divided equally between two spatial modes. Using temporal modes
rather than spatial modes gives the distinct advantage that only
two APDs are needed no matter how many times the pulse is split.
This is a significant improvement over the $N$ APDs that are
necessary in early detector array schemes~\cite{cascade, cascade2,
cascade3, cascade4} though there is a trade-off between the number
of modes and the detection rate of the detector.

\section{Experimental setup}
\subsection{Detector construction}
\begin{figure}\centerline{\scalebox{0.70}{\includegraphics{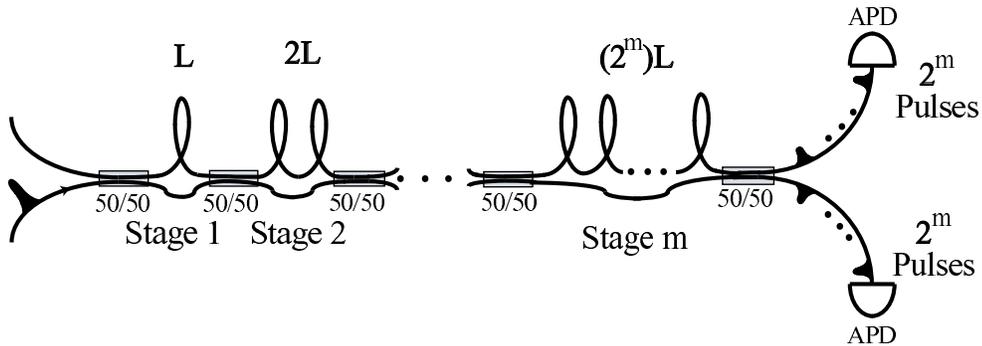}}}
\caption{Schematic setup of the detector; 50/50: symmetric fiber
couplers, APD: avalanche photodiode.} \label{fig:scheme}
\end{figure}

The general scheme for the fibre-assisted time-multiplexed
detector (TMD) is shown in Fig.~\ref{fig:scheme}.  It consists of
various lengths of single-mode optical fibre and symmetric $2
\times 2$ fibre couplers.  Light pulses propagating in a
single-mode fibre are incident upon one input of a $50/50$ coupler
whilst the other input has no incident light. The pulse is split
into two modes at the coupler where one mode is a fibre of
negligible length and the other has a much longer length $L$. This
coupler can be described by the operator transformation
\begin{equation}\label{eqn:coupler}
\hat{a}^\dagger \to (\hat{a}_s^\dagger +
\hat{a}_l^\dagger)/\sqrt{2}
\end{equation}
where the operators $\hat{a}_s^\dagger $ and $\hat{a}_l^\dagger$
are the creation operators for photons in the short and long
fibre, respectively.  This first pair of short and long fibres
along with the $50/50$ coupler that recombines them will be
referred to as the first stage of the TMD.  After the
recombination at this coupler there will be four modes: two
temporal modes in each of the two spatial modes.  The length $L$
of the fibre creates a delay between the pulses, $\Delta t =
nL/c$, where $n$ is the group index of the fibre and $c$ is the
speed of light in vacuum. This time delay is chosen to be
substantially longer than the deadtimes $t_d$ of the APDs. The two
commercial silicon APD single photon detectors (Perkin Elmer
SPCM-AQR-13) used had estimated deadtimes of 50--60 ns. Therefore
the length $L$ was chosen to be approximately 25 meters, making
$\Delta t \approx 125 $ ns. The length and the deadtimes of the
APDs determines the maximum duration of the pulses that can be
measured by the TMD. If a pulse were longer than the $\Delta t
-t_d$, the APDs would not always have time to recover from the
detection event from one mode before the light from the subsequent
temporal mode arrived at the APD.

Generally, any number of stages may be concatenated together,
where each successive stage has a fibre of length $(2^s)L$, where
$s$ refers to the stage number. The second fibre that connects the
couplers is too short to be of consequence and all the couplers
can be described by transformations similar to
Eqn.~\ref{eqn:coupler}.  If $m$ stages are used, then the pulse
will be distributed amongst $2^{m+1}$ modes, $2^m$ in each output
fibre from the final 50/50 coupler.  The time it takes for the
light to traverse the longest path of the TMD must be larger than
the time between incident pulses in order to prevent signals from
two different pulses being in the fibre at the same time. Although
increasing the number of stages allows one to detect higher photon
numbers, the time separation $\Delta t$ between temporal modes
limits the effective response time of the TMD. Therefore the
maximum repetition rate possible is approximately $(3 \times 125
\textrm{ ns})^{-1} \approx 2.5 \textrm{ MHz}$ for the two stage
TMD and about $1$ MHz for the three stage detector.

Two TMDs were constructed and tested; a two stage device~\cite{US}
and a three stage device~\cite{THEM}.  The two stage TMD was
fabricated with single mode optical fibers at 780 nm (Lucent
SMC-AO780B). A standard laser diode pulsed at 777nm was used as
the light source in the two stage TMD experiments. The pulse
duration was approximately 14 ns and the repetition rate was 10
kHz. The three stage TMD was also built from single mode fiber
supporting visible light propagation (610 nm to 730 nm) (3M
FS-SN-3224). The light source in these experiments was a laser
diode at 680 nm with 50 ps pulse duration and 20 kHz repetition
rate.  The choice of wavelength was based on a balance between
detector efficiencies and fibre losses.  The losses in fibres at
telecom wavelengths are lower than in the near infrared, but APDs
are less efficient at such wavelengths and have higher dark count
rates.

\subsection{Collecting data}

%

\begin{figure}\centerline{\scalebox{0.7}{\includegraphics{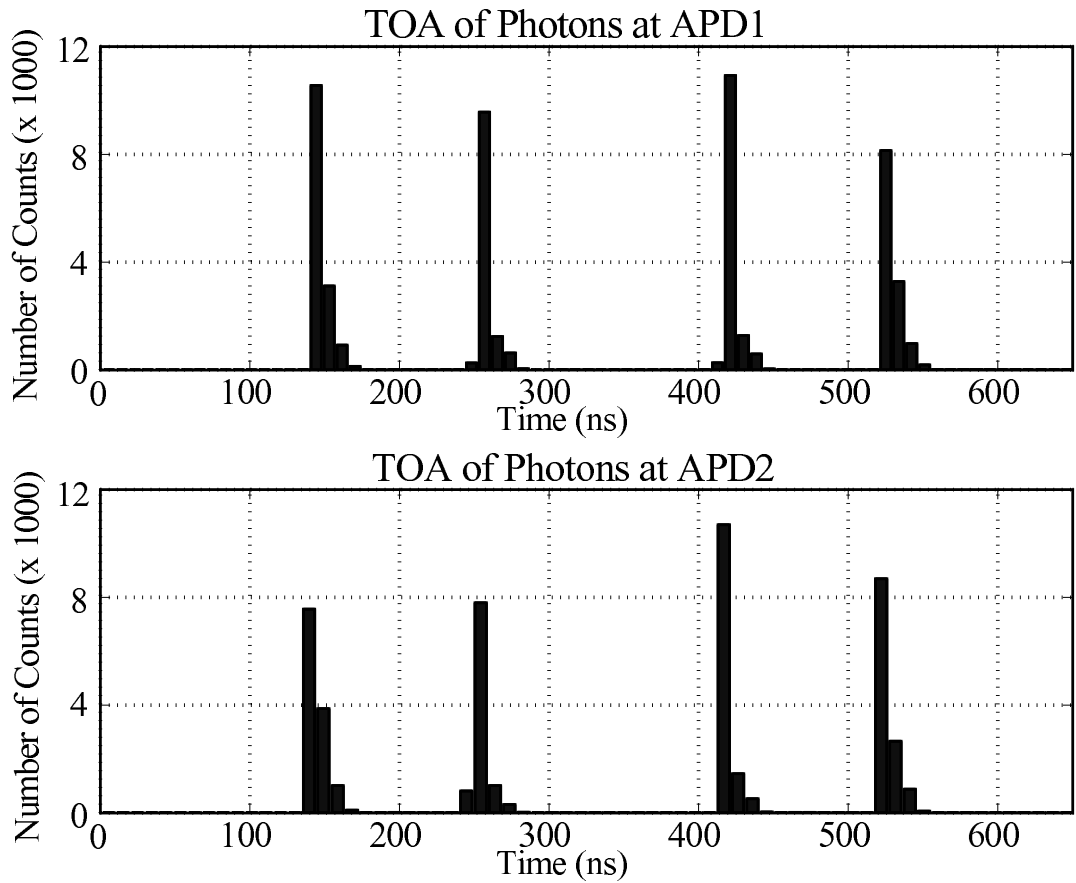}}}
\caption{The time of arrival (TOA) of counts recorded by the APDs
of the TMD.} \label{fig:toa}
\end{figure}

Data collection for the two stage TMD was performed using a
digital oscilloscope.  The oscilloscope was controlled by a PC and
used the same trigger as the laser diode that created the light
pulses. The intensity of the laser diode was attenuated with
neutral density filters and the photon distributions were
reconstructed from $10^4$ laser pulses.  The key information
extracted from single-shot measurements was the number of pulses
counted by the two APDs and their times of arrival (TOA) after the
trigger pulse. Plotting the counts against the TOA, we can clearly
identify eight temporal modes (see Fig.~\ref{fig:toa}), which are
well separated from one another. This data allows us to perform
two tasks. First, by integrating the counts in each mode we can
determine the probability that a single photon will end up in that
mode. Ideally the probability for a photon to end up in each mode
would be equal, but due to losses in the different lengths of
fibre and imperfect couplers, the probabilities differ slightly
from $2^{-(m+1)}$.

Secondly, using the TOA plot we can apply gating to the signal
from the APDs. This will reduce false counts from dark counts and
afterpulsing. These false counts could be neglected in our
experiment (they are not visible at the scale of
Fig.~\ref{fig:toa}) except for coherent pulses that have a
non-zero probability of detecting photon numbers equal to the
number of modes, $N$. In this case gating can play an important
role (see discussion below).

Afterpulsing is caused by the release of a charge that was caught
in the depletion layer of the APD in the course of the previous
avalanche.  Detection events arising from afterpulsing usually
occur as soon as the bias voltage across the diode returns to the
breakdown voltage, which means that the time between an event
triggered by a photon and the afterpulse will be on the order of
the deadtime of the detector. Therefore we expect afterpulsing
events in a time window of about 50--60 ns after each temporal
mode. Determining the afterpulse frequency involved counting the
coincidences in a single shot measurement of an event within the
specified temporal mode of the TMD with an event occurring 50--60
ns after the time window of the temporal mode. The event detected
outside the allowed time window of the temporal mode was then
labelled as an afterpulsing event.  It was found that
approximately $60\%$ of the events outside the gate window were
due to afterpulsing.  Accepting pulses only within a specified
time window allowed us to reduce the effects of both afterpulsing
and dark counts.

The three stage detector counted photon statistics using a
custom-made electronic circuit that only allowed detection in a
specified time window after each trigger~\cite{THEM}.

\section{Theory and data analysis}
\subsection{Conditional probabilities}
In general, the count statistics that are recorded by the APDs do
not represent the photon statistics of the incident pulses. For
the moment, we will assume that losses can be represented by a
beam splitter before the TMD (see section~\ref{sect:loss} for
further discussion). With this assumption, the counting statistics
are related to the incoming photon number distribution by
\begin{equation}\label{eqn:p(k)}
p_k = \sum_{n} p(k|n) \varrho_n,
\end{equation}
where $p_k$ is the probability of detecting $k$ counts,
$\varrho_n$ is the probability that there was $n$ photons incident
on the TMD, and $p(k|n)$ is the conditional probability that $k$
counts are detected when $n$ photons are incident on the TMD.
Eqn.~\ref{eqn:p(k)} can be represented as a matrix equation and
from this point forward, all equations will be written in matrix
form.  For this purpose, we associate elements of the conditional
probability matrix $\bf{C}$ with the conditional probabilities
$p(k|n)$ by the equation $c_{kn}=p(k|n)$, and the elements of
$\bm{p}$ and $\bm{\varrho}$ are $p_k$ and$\varrho_n$,
respectively. In matrix notation, Eqn.~\ref{eqn:p(k)} becomes
\begin{equation}\label{eqn:Matrix_p(k)}
\bm{p} = \bf{C} \cdot \bm{\varrho}.
\end{equation}

The conditional probabilities can be calculated from a simple
stochastic model. The problem reduces to the archetypal
probability theory example of taking $n$ balls, randomly
distributing them amongst $N$ bins and calculating the probability
that $k$ bins are occupied.   Since this is a photon counting
experiment, the photons may be treated as balls.  In this case,
$N$ is the number of temporal modes, $n$ are the number of
incident photons, and $k$ corresponds to the number of counts
recorded in the APDs.  A few results are given below in order to
illustrate how the conditional probabilities were calculated:
\begin{itemize}
\item {\bf Case 1:} $k>n$ or $n=0$ -- Assuming we can ignore dark
counts and afterpulsing, it is obvious that we will never detect
more counts in the APDs than the number of incident photons.
Therefore $p(k|n) = 0$ for $k>n$ and for $n=0$.

\item {\bf Case 2:} $n=1$ -- If we distribute a single photon
amongst $N$ modes, one and only one mode {\em must} be occupied
since the losses are being ignored; therefore $p(k|n) = 1$.

\item {\bf Case 3:} $n=2$ -- Distributing two photons into $N$
modes can only result in either one or two modes being occupied.
The probability that two modes are occupied, i.e. we detect two
counts, is then, $p(2|2) = \sum_{i \ne j} p_i p_j$ and the
probability that only one mode is filled, i.e. that both balls are
in the same mode, is $p(1|2) = \sum_i p_i^2$, where $p_i$ is the
probability that a photon will end up in mode $i$.

\item {\bf Case 4:} $p(4|7)$ -- This a specific example of one of
the more complex conditional probabilities.  In this case the APDs
count four photons when seven photons were incident on the TMD.
The equation that describes this probability is:
\begin{equation}\label{eqn:prob}
p(4|7) =  \sum_{i\ne j\ne k\ne l} \left[\binom{7}{4} p_i^4 p_j p_k
p_l + \binom{7}{3}\binom{4}{2} p_i^3 p_j^2 p_k p_l +
\frac{1}{3!}\binom{7}{2}\binom{5}{2}\binom{3}{2}p_i^2 p_j^2 p_k^2
p_l\right].
\end{equation}
The binomial coefficients account for the different ways the
photons can be distributed amongst the modes and the factorial in
the denominator corrects for the overcounting due to having three
modes occupied by the same number of photons.
\end{itemize}

Note that this approach is general enough to apply to a TMD with
any number of temporal modes $N$ and allow each temporal mode to
have its own unique probability of containing a photon.  These
different probabilities were obtained from the integration of the
temporal modes in Fig.~\ref{fig:toa} and resulted in the
probabilities: (0.141, 0.112, 0.125 0.121, 0.132, 0.105, 0.134,
0.129), which are all close to $1/8$. The conditional
probabilities were worked out for several different mode weights
and it was shown that small deviations from $1/8$ do not
dramatically affect the $p(k|n)$.  Still, the different weights
were kept in the analysis to obtain the best accuracy.

The conditional probability matrix is a $9\times 9$ matrix. We
include the $k,n = 0$ terms, as they become important when we
discuss losses (see Sect.~\ref{sect:loss}). For the eight weights
of the temporal modes given above, the result is:
\begin{equation}\label{eqn:P_kn}
\bf{C} = \left[ \begin{array}{ccccccccc}
1&0&0&0&0&0&0&0&0\\\noalign{\medskip}0& 1& 0.126& 0.016& 0.002& 0&
0& 0& 0\\\noalign{\medskip}0&0& 0.875& 0.330& 0.097& 0.026& 0.007&
0.002& 0\\\noalign{\medskip}0&0&0& 0.655& 0.494& 0.260& 0.118&
0.050& 0.020\\\noalign{\medskip}0&0&0&0& 0.408& 0.512& 0.420&
0.285& 0.175\\\noalign{\medskip}0&0&0&0&0& 0.203& 0.383& 0.449&
0.423\\\noalign{\medskip}0&0&0&0&0&0& 0.076& 0.200&
0.317\\\noalign{\medskip}0&0&0&0&0&0&0& 0.019&
0.066\\\noalign{\medskip}0&0&0&0&0&0&0&0& 0.002 \end{array}
\right].
\end{equation}
The procedure for obtaining the photon number distribution from
the count statistics using this matrix and
Eqn.~\ref{eqn:Matrix_p(k)} is described in the next section.

\subsection{Photon number reconstruction}
In our experiments, we take data over a large number of laser
pulses and reconstruct the photon number distribution from the
count statistics of the APDs.  Because we use a coherent light
source, the photon number distribution is Poissonian :
\begin{equation}\label{eqn:poisson}
P(\mu, n) = \frac{\mu^n e^{-\mu}}{n!},
\end{equation}
where $\mu$ is the mean photon number and $n$ is the number of
photons.  In general losses modify the photon distribution of a
quantum state.  Such is the case with Fock states and squeezed
states.  However, for a coherent state losses simply reduce the
mean of the distribution, leaving the higher moments of the
probability distribution as they were.  We therefore introduce
$\mu' = \eta\mu$, where $\eta$ takes into account both the lossy
fibers and the inefficiency of the detectors.  Several approaches
can be taken to determine $\bm{\varrho}$ from $\bm{p}$ and each
method implies a different set of assumptions.  Our first two
methods assume that the distribution will be Poissonian, but the
third is more general, as it assumes nothing about the form of the
distribution.

The first method assumes that each temporal mode of the TMD is
equally likely to contain a photon.  This implies the different
length fibres do not affect the losses in anyway and that the
$50/50$ couplers are perfect.  By using $\mu'$ instead of $\mu$ in
Eqn.~\ref{eqn:poisson}, we see that the probability of detecting
no photons is $P_0 = \exp (-\mu')$ and the probability of
detecting at least one photon is $P_A = 1-P_0$. Since the input
state is assumed to be coherent, there exist no correlations
between the measured counts and the detection of photons in each
of the temporal modes is independent. Hence the probability of
obtaining exactly $k$ detection events is given by the binomial
distribution
\begin{equation}\label{eqn:binom}
p_k = \frac{N!}{(N - k)! \, k!}(P_0)^{N-k}(1-P_0)^k,
\end{equation}
where $N$ is again the number of temporal modes in the TMD.

%

\begin{figure}
\includegraphics*[width=0.7\textwidth]{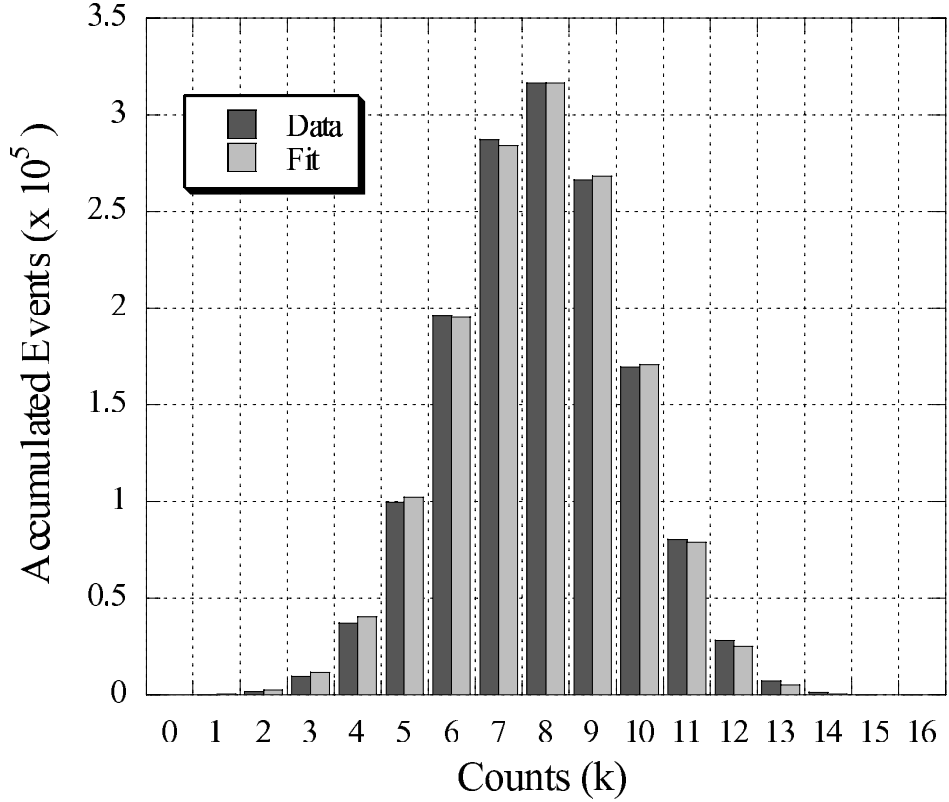}%
\caption{Histogram of the number of events in which $k$ photons
were detected for an incident coherent state pulse with a
relatively large intensity. The light bars correspond to the
theoretical prediction of Eqn.~\ref{eqn:binom} based on a
least-squares fit with $\mu'=13.1$, while the dark bars correspond
the experimental\ measurements. The errors are the square root of
the number of events in each column. \label{fig:mean13}}
\end{figure}

Fitting experimental data from the three stage TMD to
Eqn.~\ref{eqn:binom} with the mean photon number $\mu'$ and a
normalisation constant as fit parameters, reveals how well this
simple theoretical model works. Experimental data for $\mu' =
13.1$ is shown in Fig.~\ref{fig:mean13}.  The deviations of the
experiment from the theory are likely to be from assuming equal
weighting of each temporal mode.

%

\begin{figure}
\includegraphics*[width=0.55\textwidth]{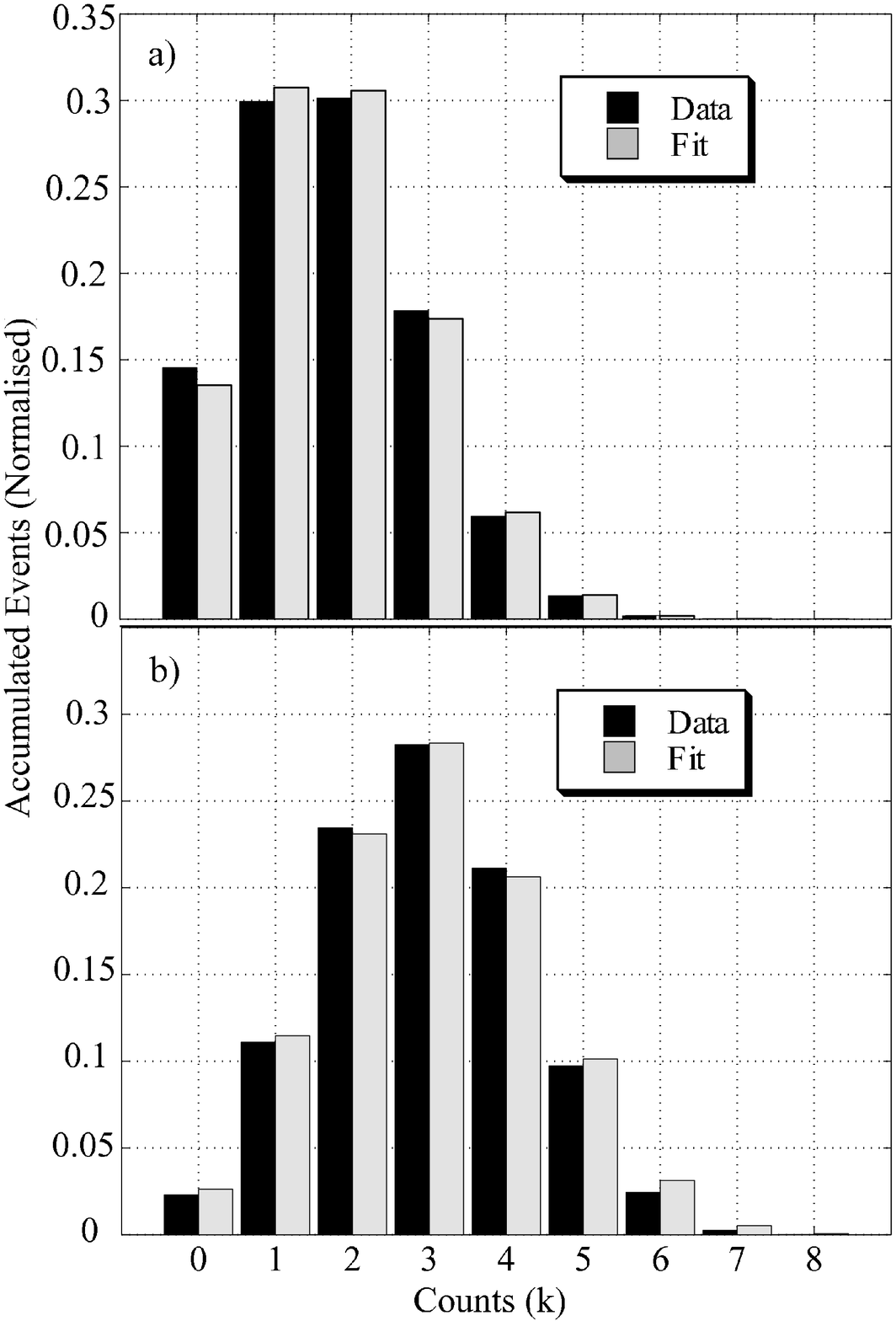}%
\caption{Comparison of the theoretical prediction and experimental
results for a weak coherent state input by fitting the data to a
coherent state modified by the conditional probability matrix.
Here the best fit corresponds to (a) $\mu'=2.00$ and (b)
$\mu'=3.77$. \label{fig:OX_iter}}
\end{figure}

The second method of determining the incoming photon distribution
employs the conditional probability matrix $\bf{C}$ in conjunction
with Eqn.~\ref{eqn:Matrix_p(k)} and uses a least-squares routine
to fit the data to a Poissonian distribution.  This is
accomplished by assuming a Poissonian distribution of different
mean value for $\bm{\varrho}$, acting on it with $\bf{C}$, and
comparing the resulting $\bm{p}$ with the data.  The mean photon
number of the light pulses are obtained from the modified
Poissonian distribution that varied least with the experimental
data. Experimental count statistics from the two stage TMD are
shown with their associated fits for $\mu' = 2.00$ and $\mu' =
3.77$ in Fig.~\ref{fig:OX_iter}a and \ref{fig:OX_iter}b,
respectively.
%

\begin{figure}
\includegraphics*[width=0.9\textwidth]{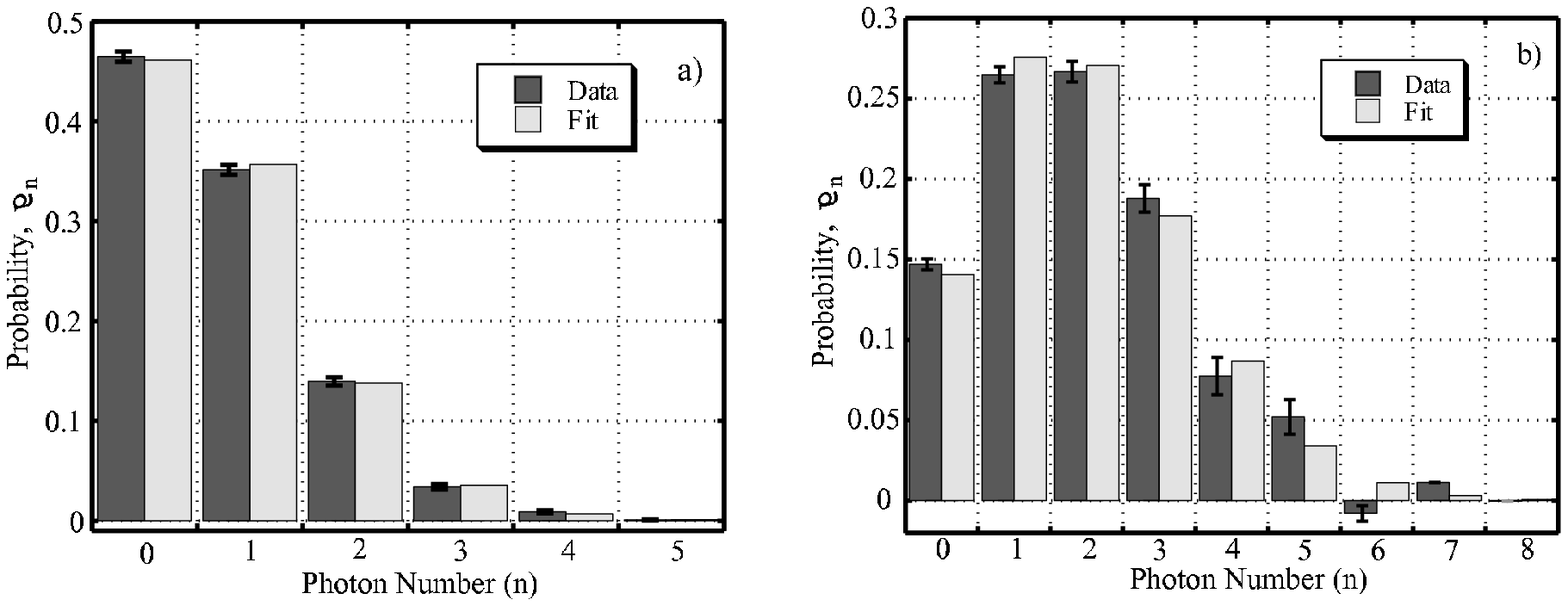}%
\caption{Photon number distribution of a coherent state recreated
by the inversion method for a state with (a) $\mu' = 0.77$ and (b)
$\mu' = 2.00$.  A mean photon number of two is the highest value
that the inversion gives a reasonable result and slight negative
probabilities begin appearing.
 \label{fig:Iter_Err}}
\end{figure}

\begin{figure}
\includegraphics*[width=0.5\textwidth]{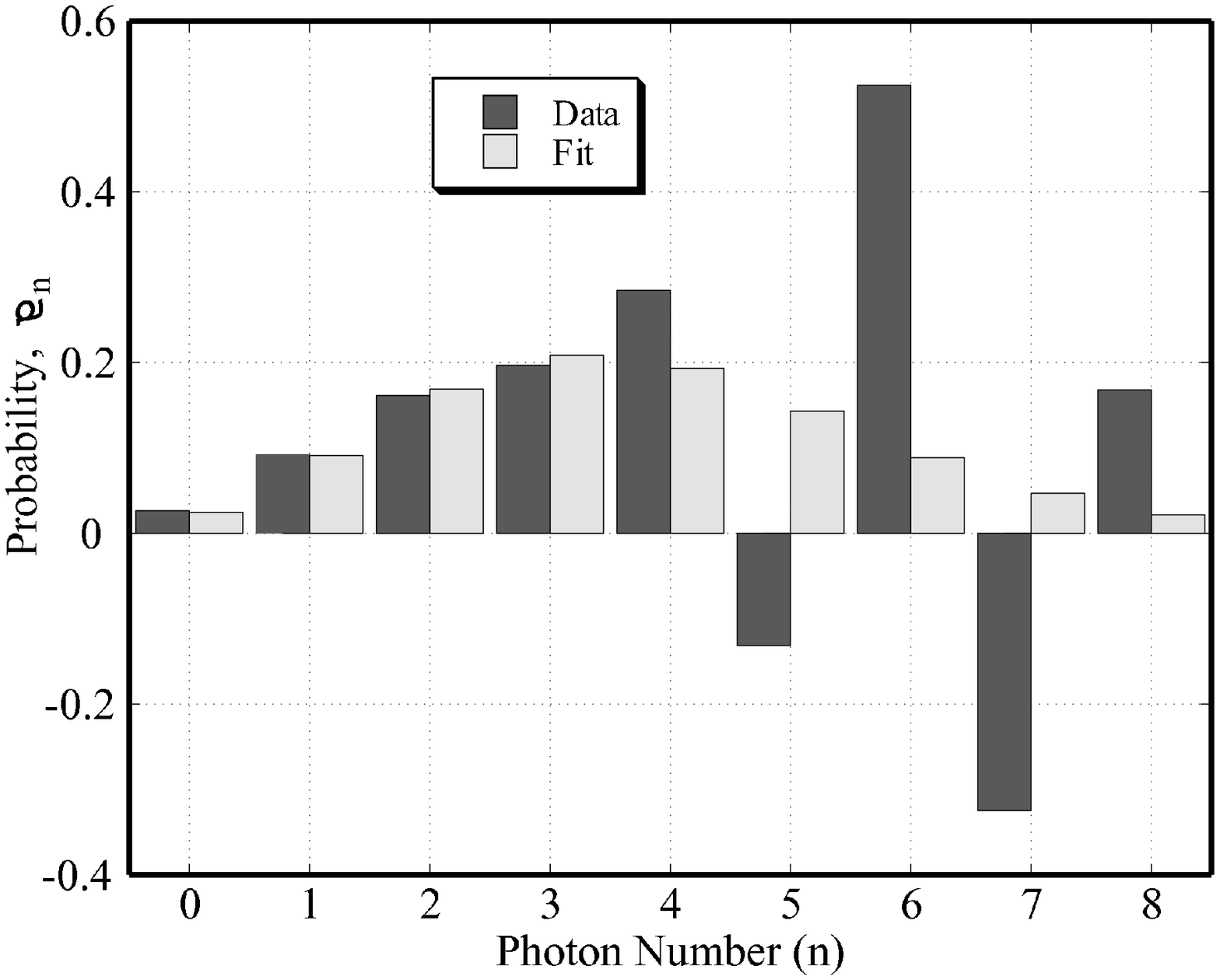}%
\caption{Demonstration of the problems that arise when using the
inversion method for a coherent state with too large of a mean
photon number.  Here the mean value of the Poissonian distribution
is $\mu'=3.77$. \label{fig:inv_singular}}
\end{figure}

The third method of reconstructing the photon statistics makes no
assumptions, not even about the form of the photon statistics.  It
is the most general way to determine the input number
distribution, but is less robust than the previous two methods.
Inverting Eqn.~\ref{eqn:Matrix_p(k)}, we obtain:
\begin{equation}\label{eqn:rho(n)}
\bm{\varrho} = \bf{C}^{\mathnormal{-1}} \cdot \bm{p}.
\end{equation}
This inversion breaks down if there are many events with $k
\approx N$. Inspection of the conditional probability matrix
reveals the reason for this failure: $\bf{C}$ is an upper
triangular matrix with the lower-right diagonal elements having
very small values. Thus the matrix is singular and cannot be
inverted. If the data contains events where $p_k \ne 0$ for a
value of $k$ corresponding to one of these large elements, the
reconstructed photon number distribution will acquire unphysical
properties such as negative probabilities and probabilities
greater than one.  It is important to re-emphasize that these
unphysical results are a result of the instability of the
inversion. Experimental data was inverted to give the photon
probability distributions for three different coherent states: one
possesses an easily invertible photon number ($\mu'=0.77$), the
second has a mean that is on the border of being invertible and
being singular ($\mu'=2.00$), and the last is a state well above
the point of inversion failure ($\mu'=3.77$)(see
Fig.~\ref{fig:Iter_Err}a, \ref{fig:Iter_Err}b, and
\ref{fig:inv_singular} respectively). The error bars in
Fig.~\ref{fig:Iter_Err} were determined from 1000 Monte Carlo
simulations of the experiment with the appropriate number of
pulses ($10^4$).  Error bars are not shown in
Fig.~\ref{fig:inv_singular} because the inversion is too unstable
for error bars to be meaningful.

%
%

It should be noted that this inversion process can be improved, as
we put no constraints on the inversion.  It is certainly possible
to force the inversion algorithm to exclude unphysical results
such as negative probabilities or probabilities greater than one.
If these constraints were implemented and a maximum likelihood
technique~\cite{Max-Like} was used, then the inversion would most
likely work both more accurately and for higher photon numbers.

\subsection{Losses}\label{sect:loss}
In the previous analysis losses were ignored because they did not
affect the distributions of the coherent states that were used in
the experiment.  However, as previously stated, for nonclassical
states the efficiency of the detector is crucial.  The largest
sources of loss arise from coupling into the fibre, absorption and
scattering of light inside the fibre, and the imperfect
efficiencies of the APDs.  The absorptive losses depend on the
amount of fibre through which the light travels.  For a fibre of
length $L$, the fraction of the incident power that is transmitted
will be $f$.  The transmission through a fibre of length $2L$ is
then $f^2$, etc.

\begin{figure}
\includegraphics*[width=0.9\textwidth]{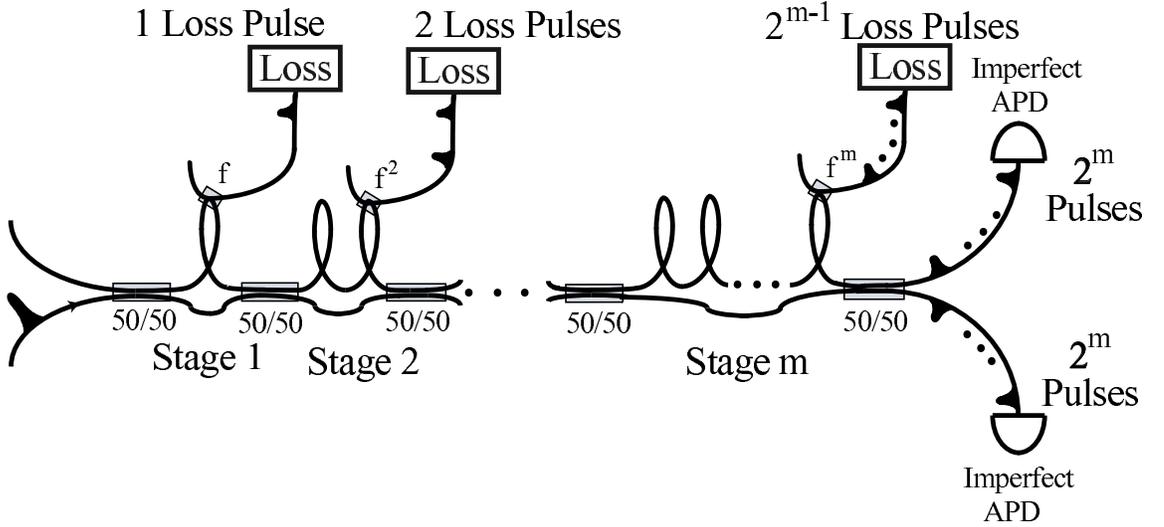}%
\caption{A model for the losses based on attributing the losses to
fibre absorption and scattering.  Loss in the fibre of length $L$
is modelled by a coupler that transmits a fraction $f$ of the
light and reflects $1-f$ of the light into a loss channel.  The
extra modes can be treated the same way as the regular temporal
modes of the TMD.   \label{fig:APL_Loss}}
\end{figure}

One way to simulate the effect of the lossy fibre (used for the
three stage TMD) is to model it as a section of lossless fibre
($f=1$) followed by a coupler in each stage of the TMD (see
Fig.~\ref{fig:APL_Loss}). These additional couplers transmit a
fraction of the light $f$ and reflect $1-f$ of the light into a
different fibre.  It is assumed that the short sections of fibre
connecting the $50/50$ couplers have negligible losses. This
creates $\sum_{s=0}^{m-1} 2^s$ extra loss modes, each with a
different probability of being occupied by a photon.  These extra
modes can be treated in the same manner as the regular temporal
modes of the TMD except that the photons of the lossy modes will
not be detected.  After all the transformations (see
Eqn.~\ref{eqn:coupler}) for both the actual couplers and the loss
couplers are applied, the input state can be written as a large
number of terms that correspond to all the ways that the $n$
incident photons can be distributed over all 23 modes, for the
three-stage TMD.

In order to account for the nonunit efficiency of the APDs, we
will assume that $P_0 = (1-\eta)^{n'}$, where $P_0$ is the
probability that the APD detects no photons when $n'$ photons are
incident upon it.  The probability of detecting at least one
photon is then $P_A = 1-P_0$.  The value of $P_0$, and hence
$P_A$, were determined experimentally and the probability of
obtaining $k$ detection events was calculated.  This contribution
was then added to the probability distribution $p_k$ for each term
of the state.

\begin{figure}
\includegraphics*[width=0.45\textwidth]{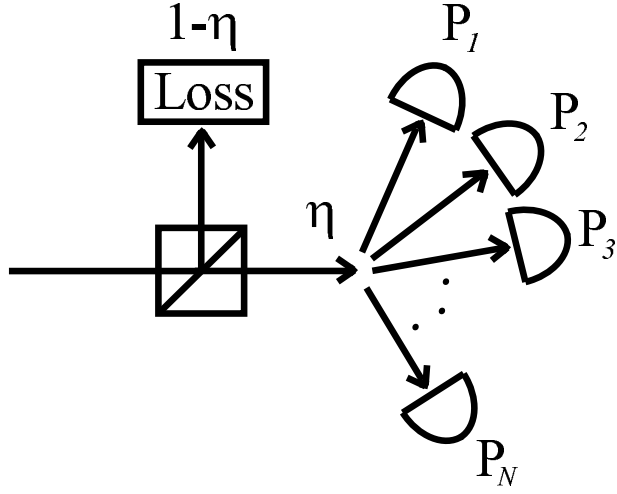}%
\caption{A model for the losses where the physical source of the
losses are ignored.  For this model it is assumed that all the
loss happens prior to the TMD and then the optical pulse is split
into $N$ channels with different weights $P_i$ which are
determined experimentally. \label{fig:OX_Loss}}
\end{figure}

This method of modelling the losses focussed on the experimental
setup and the actual sources of the loss.  We can, however, think
about the losses, and the experiment in general, in a more
abstract manner.  By doing so we allow for a much more general
method of taking losses into account.  The first aspect changed to
make the scheme more abstract will be to simply think of the TMD
as a device that splits a pulse into $N$ modes (see
Fig.~\ref{fig:OX_Loss}).  The other input modes of this multi-port
beam splitter have been ignored as they are left in the vacuum
state and therefore play no role in the model.  We don't care how
the modes were created nor do we care what type of modes they are
(i. e. temporal or spatial). These modes only have a few qualities
that are of importance:
\begin{enumerate}
\item Each mode is easily distinguishable from the others due to
the orthogonality of the modes.

\item There are no losses other than the initial beam splitter.

\item In general, each mode will have a different probability
~$P_i$ of containing a photon.

\end{enumerate}

Loss is introduced into the system by placing a single beam
splitter before the multi-port beam splitter. This beam splitter
reflects a fraction, $1-\eta$ of the incident light. Since this is
the only source of loss in the scheme, $\sum_{i=1}^{N}P_i = \eta$.
It can be shown that by choosing appropriate values of $P_i$ and
$\eta$, the losses from the method shown in
Fig.~\ref{fig:APL_Loss} can be recreated.

Another important attribute of the abstract loss model is that the
loss becomes independent of the conditional probability matrix.
This makes the analysis of how losses affect the TMD's performance
much simpler since Eqn.~\ref{eqn:Matrix_p(k)}, with losses,
becomes:
\begin{equation}\label{eqn:P_k_loss}
\bm{p} =   \bf{C \cdot L} \cdot \bm{\varrho},
\end{equation}
where $\bf{L}$ is another upper triangular conditional probability
matrix with elements given by the binomial distribution
\begin{equation}
l_{n'n} = \binom{n}{n'}\eta^{n'}(1-\eta)^{n-n'},
\end{equation}
where $n'$ is the photon number after losses are taken into
account.  Note that Eqn.~\ref{eqn:P_k_loss} is only valid for $n
\geq n'$ and $l_{n'n} = 0$ for $n<n'$.

The conditional probability matrix $\bf{C}$ no longer acts on the
photon number distribution directly, but on the reduced
distribution. With this method, retrieving the photon number
distribution is done in precisely the same way as
before~(Eqn.~\ref{eqn:rho(n)}), except the matrix we invert is the
combination of the two matrices, $\bf{L}$ and $\bf{C}$:
\begin{equation}\label{eqn:full_analysis}
\bm{\varrho} =
\bf{L}^{\mathnormal{-1}}\cdot\bf{C}^{\mathnormal{-1}}\cdot\bm{p}.
\end{equation}
Similar methods for compensating losses in photodetection have
previously been discussed~\cite{Loss, Konrad_Loss}.

\section{Applications}

With the use of Eqn.~\ref{eqn:full_analysis} we can completely
reconstruct the photon probability distribution of an optical
pulse given a large number of single shot measurements of
identical pulses.  Reconstructing coherent state photon statistics
is a good proof of principle experiment, but is not the primary
application of the TMD.

\subsection{Reconstructing Fock States}
\begin{figure}
\includegraphics*[width=0.6\textwidth]{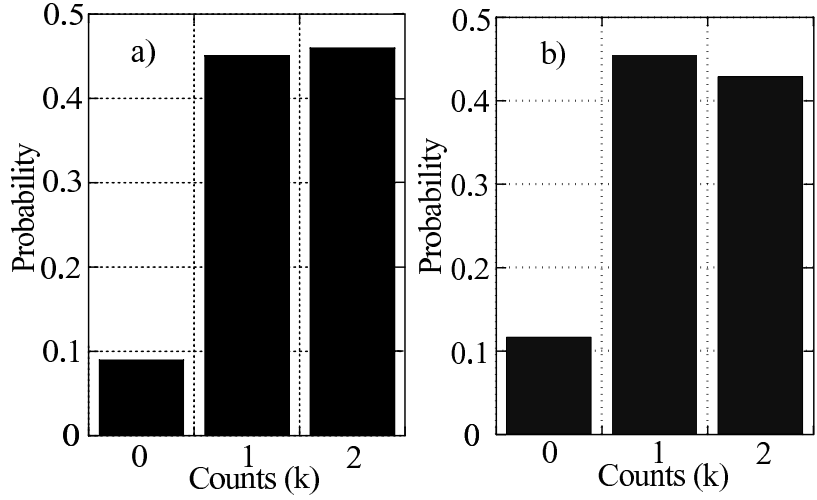}%
\caption{The calculated count statistics $p_k$ for a two photon
Fock state ($n=2$) incident on the TMD for a (a) three stage and a
(b) two stage TMD with detector efficiency $\eta=0.7$ and no other
losses ($f=1$). The three stage TMD performs better and more
stages would further improve the probability of detecting two
photons.\label{fig:Fock2}}
\end{figure}

Thus far we have only considered coherent states of light, where
losses were less important; it is crucial to minimise losses for
Fock states. Schemes for linear optics quantum computing are
concerned with distinguishing the presence of one or two
photons~\cite{KLM, LOFranson, LOQC1, LOQC2, LOQC3} and conditional
state preparation depends on being able to measure three photon
states.  To illustrate the effects losses have on a Fock state
measurement, Fig.~\ref{fig:Fock2} shows \mbox{$p_k = p(k| n=2)$}
for a detector with $70\%$ efficiency and all other losses set to
zero. Fig.~\ref{fig:Fock2}a (b) is the result for the three (two)
stage TMD.  Notice that more stages could further improve the
capabilities of the apparatus, but eventually the losses and
lowered detection rate from adding more stages would render the
detector less useful.

\begin{figure}
\includegraphics*[width=0.6\textwidth]{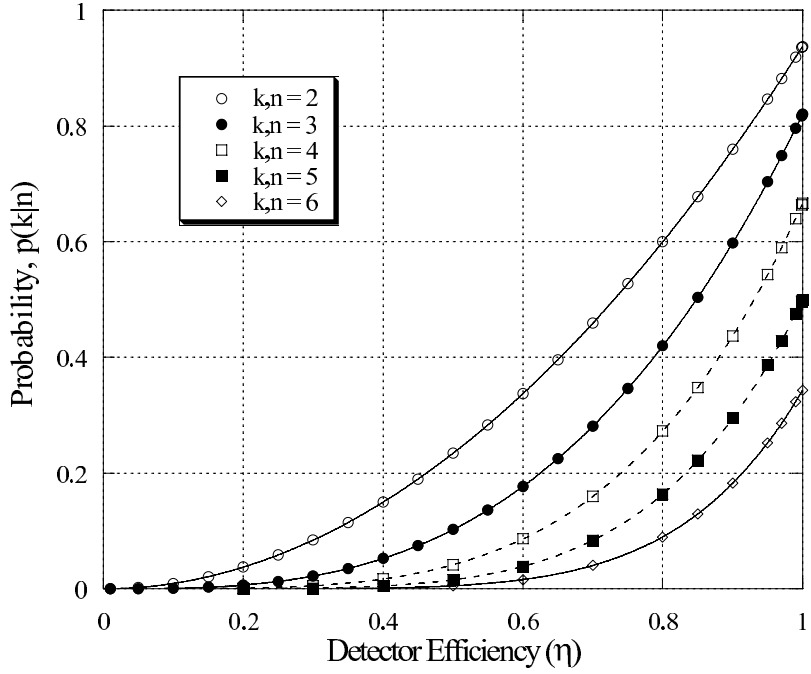}%
\caption{The calculated conditional probability $p(k=j|n=j)$ of
detecting all of the incident photons $n$ as a function of the
detection efficiency $\eta$.  The points are shown for the three
stage TMD and were calculated numerically using the operator
transformation technique of Eqn.~\ref{eqn:coupler}, while the
lines correspond to the analytic formula of
Eqn.~\ref{eqn:P_kn_APL}. \label{fig:P_kn_efficiency}}
\end{figure}

The effects of losses are further demonstrated in
Fig.~\ref{fig:P_kn_efficiency} by looking at how the
\mbox{$p(k=j|n=j)$} of the three stage TMD varies for different
detector efficiencies. All other losses being negligible and
assuming perfect $50/50$ couplers, the analytic formula for the
diagonal elements of the conditional probability matrix is
\begin{equation}\label{eqn:P_kn_APL}
p(k=j|n=j) = \frac{16!}{16^j(16-j)!}\eta^j \qquad \textrm{for}
\quad j \leq 16.
\end{equation}
This formula also shows the effect of having a finite number of
modes in the TMD. As the number of incident photons increases, the
probability of detecting all of them decreases, even for $100\%$
detector efficiency.  This is due to the increased probability
that two of the photons will be in the same mode.

\subsection{Single shot measurement}

Most of our discussion thus far has revolved around the
conditional probability $p(k|n)$ of detecting $k$ photons if $n$
photons were actually incident on the TMD.  It is interesting to
look at the inverse conditional probability, $\tilde p(n|k)$ that
there are $n$ photons in a pulse if $k$ counts were measured. The
tilde indicates that this is a different function of $n$ and $k$
than the forward probability $p(k|n)$.  $\tilde p(n|k)$ contains
information about how certain one can be about the input photon
number from a single shot measurement using the TMD. Such
measurements are the main application of the TMD and $\tilde
p(n|k)$ needs to be well understood if one would like to perform
conditional state preparation.

Using Bayes' theorem, $p(A|B) = p(A \cap B)/p(B)$, the probability
$\tilde p(n|k)$ can be written as
\begin{equation}\label{eqn:P_nk}
\tilde p(n|k) = \frac{\varrho_n}{p_k} p'(k|n).
\end{equation}
Unlike $p(k|n)$, which only depends on the construction of the
TMD, this probability is a function of the input photon
distribution as well.  The $p'(k|n)$ in Eqn.~\ref{eqn:P_nk} is the
matrix element that results from combining the conditional
probability matrix and the loss matrix.  This matrix transforms
the incoming photon number into the count distribution.
Substituting $p'(k|n) =[\bf{C\cdot L}]_{\mathnormal{kn}}$ (the
element in the $k$th row and $n$th column of $\bf{C\cdot L}$) and
using Eqn.~\ref{eqn:P_k_loss} to rewrite $p_k$,
Eqn.~\ref{eqn:P_nk} becomes
\begin{equation}\label{eqn:P_nk_messy}
\tilde p(n|k) = {\varrho_n \times [\bf{C\cdot
L}]_{\mathnormal{kn}} \over [\bf{C\cdot
L}\cdot\bm{\varrho}]_{\mathnormal{k}}},
\end{equation}
where everything is now in terms of the input distribution and the
conditional probability matrices.

The conditional probability $\tilde p(n|k)$ is a useful way to
evaluate the performance of the TMD.  It depends critically,
however, on the a priori input photon distribution used in the
analysis. For example, the Fock state with photon number $j$ has a
photon number distribution vector that is $\varrho_n =
\delta_{nj}$, meaning that the probability of there being a photon
incident on the detector is zero unless $n=j$, in which case the
probability is one. If we put this into Eqn.~\ref{eqn:P_nk_messy}
for $\varrho_n$, the result is zeros for all $\tilde p(n|k)$
except for $k \le n$, where the value is unity.  This indicates
that no matter what number of photons are detected, we are
confident that we measured the Fock state $|j \rangle$.  This
seems like a fruitless analysis at first, but the reason for the
trivial result is that one must assume the photon probability
distribution is already known.  In this case, we assumed the state
was in a specific Fock state. Therefore no matter what the result
of the photon number measurement, the answer will always tell us
that we are confident we measured $|j \rangle$.

\begin{figure}
\includegraphics*[width=0.6\textwidth]{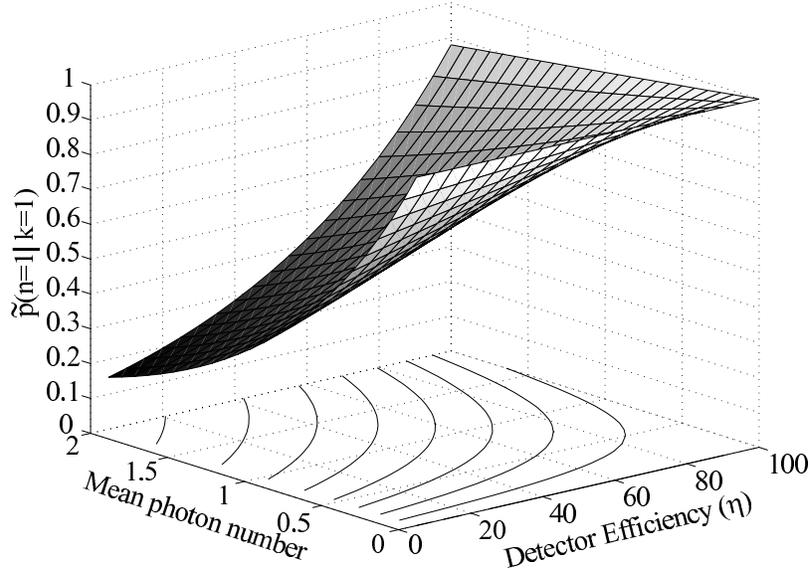}%
\caption{The calculated conditional probability $\tilde
p(n=1|k=1)$ that if one photon was detected that there was
actually exactly one photon in the pulse for coherent states with
different means and for different detector efficiencies.  Note
that the probability does not go to zero as the detector
efficiency goes to zero.
 \label{fig:P_nk_3d}}
\end{figure}

The problem arises from the fact that the TMD is performing a
projective measurement in the photon number basis and the Fock
state is an eigenstate of this projection operator.  If we use
states that are not eigenstates of the measurement, such as a
coherent state or a mixture of Fock states which we wish to
distinguish, the result is nontrivial and information is gained
from the measurement. Fig.~\ref{fig:P_nk_3d} shows a plot of
$\tilde p(n=1|k=1)$ versus detector efficiency and mean photon
number of the coherent state and essentially characterises the
performance of a single-shot measurement of a coherent state using
a two stage TMD.  It is clear that either a decrease in detector
efficiency or an increase in mean photon number will degrade the
TMD performance.  For good performance low photon numbers should
be used. Notice that $\tilde p(n|k)$ does not necessarily approach
zero, which seems counterintuitive at first, especially for a
detector efficiency that approaches zero. One must remember,
however, that this conditional probability is not the probability
of detecting a photon (which {\em does} go to zero as the
efficiency decreases) but is the probability that {\em if} a
photon was detected, no matter how unlikely it may be, that there
was exactly one photon in the pulse~\footnote{In a realistic
experiment, detector dark counts will be present, but can be
suppressed with narrow time-gating. We assume in the present
analysis that dark counts are completely negligible}.

\begin{figure}
\includegraphics*[width=0.6\textwidth]{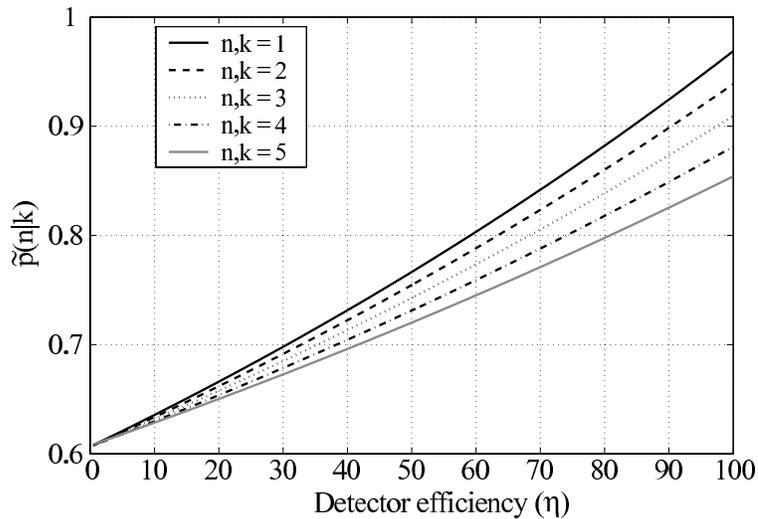}%
\caption{The conditional probability $\tilde p(n|k)$ that if $k$
photons were detected that there were actually $k$ photons in the
pulse for a coherent state with a mean photon number of $0.5$. For
$n,k=1$ this is a cross-section of Fig.~\ref{fig:P_nk_3d}.
 \label{fig:P_nk_2d}}
\end{figure}

A plot similar to Fig.~\ref{fig:P_nk_3d} can be constructed for
higher photon numbers, $n=k > 1$.  These plots all look very
similar but with lower probabilities in general. To illustrate
this Fig.~\ref{fig:P_nk_2d} shows a plot of $\tilde p(n|k)$ versus
the detector efficiency for a coherent state of mean photon number
equal to $0.5$ for different values of $n=k$. For $n,k=1$, this
plot is a cross-section of the plot in Fig.~\ref{fig:P_nk_3d}.

\section{Summary}

In summary, we have described in detail the construction and
performance of a time-multiplexed single-photon detector with
photon number resolution.  The TMD is easily constructed from
single-mode optical fibre, 50/50 couplers and two standard silicon
APDs and does not require extreme operating conditions like other
photon resolving detectors.  Both two~\cite{US} and
three~\cite{THEM} stage detectors were constructed, creating eight
and sixteen modes respectively, and the TMD is extendible to any
number of stages, limited by the losses introduced by each
additional stage.

The photon statistics of different coherent states of light were
experimentally reconstructed as a proof-of-principle experiment.
Good agreement was seen between the experiment and the theories
developed to model the action of the TMD.

An analysis of the single-shot performance of the detector shows
that the TMD can be a useful tool for conditional state
preparation and other quantum information schemes since the
conditional probability of a single count being caused by single
photon remains reasonably high even for inefficient detectors.

\section*{Acknowledgements}

We are grateful for conversations with C. Radzewicz and M. T.
Lamar. This research was supported by the US Department of Defense
through the Army Research Office via grant number DAAD19-02-1-0163
(Oxford) and ARO, NSA, ARDA, and IR\&D (Applied Physics
Laboratory). C. {\'S}liwa's current address is: Centre for
Theoretical Physics, Al. Lotnikow 32/46, 02-668 Warszawa, Poland.

\end{document}